\def\year{2020}\relax
\documentclass[letterpaper]{article} % DO NOT CHANGE THIS
\usepackage{aaai20}  % DO NOT CHANGE THIS
\usepackage{times}  % DO NOT CHANGE THIS
\usepackage{helvet} % DO NOT CHANGE THIS
\usepackage{courier}  % DO NOT CHANGE THIS
\usepackage[hyphens]{url}  % DO NOT CHANGE THIS
\usepackage{graphicx} % DO NOT CHANGE THIS
\usepackage{graphicx}  % DO NOT CHANGE THIS
\usepackage{bm}
\usepackage{amsmath,amssymb,amsfonts}

\newcommand*\rot{\rotatebox{90}}

\title{Multi-scale Anomaly Detection on Attributed Networks}
\author{Leonardo Guti\'errez-G\'omez,\textsuperscript{\rm 1,3}\thanks{Corresponding author} Alexandre Bovet,\textsuperscript{\rm 1} Jean-Charles Delvenne,\textsuperscript{\rm 1,2}\\
\textsuperscript{\rm 1}Institute for Information and Communication Technologies, Electronics and Applied Mathematics (ICTEAM) \\ \textsuperscript{\rm 2}Center for Operations Research and Econometrics (CORE) \\ Universit\'e catholique de Louvain, Louvain-la-Neuve, Belgium \\
alexandre.bovet@uclouvain.be, jean-charles.delvenne@uclouvain.be 
\\ \textsuperscript{\rm 3}Luxembourg Institute of Science and Technology (LIST), Esch-sur-Alzette, Luxembourg \\
leonardo.gutierrez@list.lu
}

\begin{document}

\maketitle

\begin{abstract}
Many social and economic systems can be represented as attributed networks encoding the relations between entities who are themselves described by different node attributes. Finding anomalies in these systems is crucial for detecting abuses such as credit card frauds, web spams or network intrusions. Intuitively, anomalous nodes are defined as nodes whose attributes differ starkly from the attributes of a certain set of nodes of reference, called the \textit{context} of the anomaly.
While some methods have proposed to spot anomalies locally, globally or within a community context, the problem remain challenging due to the multi-scale composition of real networks and the heterogeneity of node metadata. Here, we propose a principled way to uncover outlier nodes simultaneously with the context with respect to which they are anomalous, at \textit{all relevant scales} of the network.
We characterize anomalous nodes in terms of the concentration retained for each node after smoothing specific signals localized on the vertices of the graph. Besides, we introduce a graph signal processing formulation of the Markov stability framework used in community detection, in order to find the context of anomalies. The performance of our method is assessed on synthetic and real-world attributed networks and shows superior results concerning state of the art algorithms. Finally, we show the scalability of our approach in large networks employing Chebychev polynomial approximations.
\end{abstract}

\noindent 
Anomaly detection is an important problem in data mining with multiple applications in diverse domains \cite{Aggarwal:2013:OA:2436823}. Examples include credit card fraud detection, %\cite{CORREABAHNSEN2016134} 
intrusion detection for network systems, %\cite{Zhang:2005:APD:1086297.1086305,Ding:2012:ISC:2339530.2339670}, 
identifying anomalous users in communication networks, %\cite{6680745}, 
web spam detection, %\cite{Castillo:2007:KYN:1277741.1277814} 
and so on. 
Anomalous data can be understood as noteworthy objects with patterns or behaviors that deviate significantly from a background property.
Investigating anomalies in networks is especially interesting because many real world systems are well modeled by a network consisting of nodes with specific attributes and edges representing the relations between nodes.
For instance, in a social network where nodes represent persons and edges social ties, node attributes could be demographic information. In a co-purchase network, nodes represent items and might contain information about them as attributes, while an edge is present between two items when the same person has bought them.

\begin{figure}[t!]
\centering
\hspace*{-0.3cm}
\includegraphics[height=2.9cm]{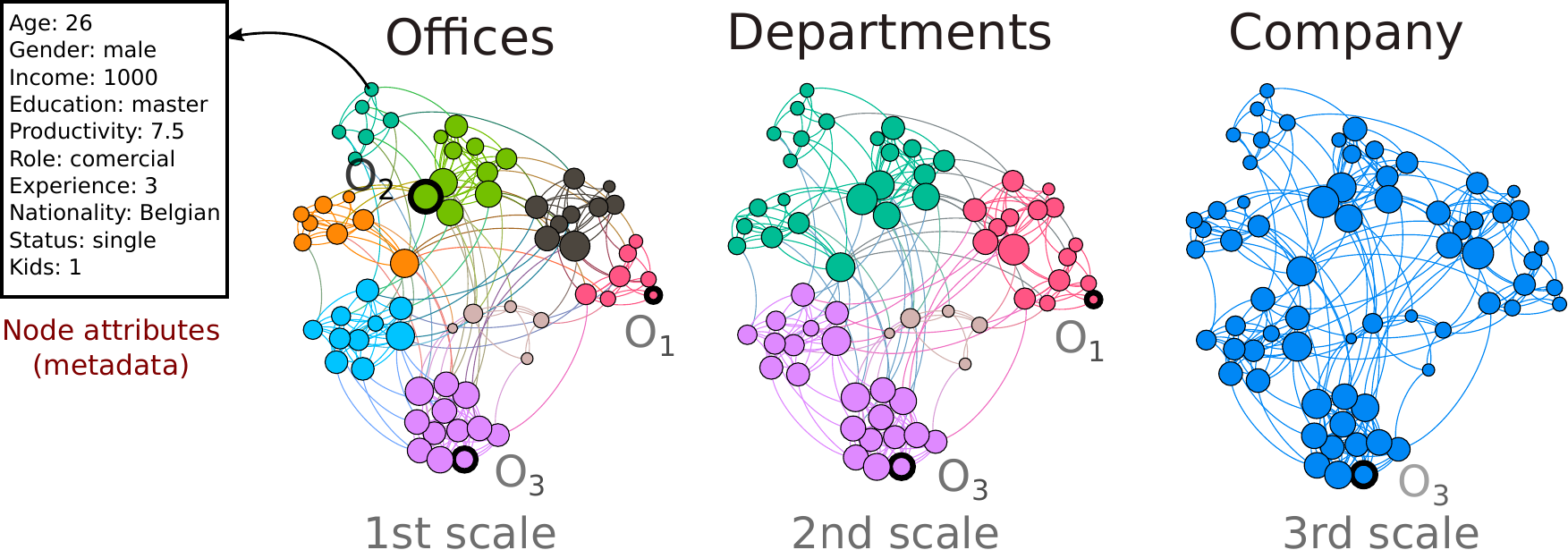}
\caption{A toy example of work relation network. Nodes have attributes describing individual features. Node attributes define structural clusters in multiple scales. At the \textit{1st scale} outlier nodes ($O_1, O_2, O_3$) lie within a local context, i.e, offices. In a \textit{2nd scale}, departments emerge as new contexts where $O_2$ is not defined. Finally, at a larger scale $O_3$ remains as a global anomaly in context of the whole company.}
\label{office_example}
\end{figure}

Because the mechanisms generating anomalies in real world networks are usually unknown, defining ground truth anomalies, is problematic and measuring the degree for which a node is anomalous is a challenging problem. For this reason, anomalies must be defined in opposition to a background of ``normal'' nodes, i.e., the \textit{context} relevant for an anomaly. 

So far, the issue of finding anomalous nodes, or outliers (we consider these words as synonyms in this paper) on networks together with their relevant contexts has not been entirely solved. Existing approaches consider outliers within local or community contexts \cite{ijcai2017-325,Gao:2010:COE:1835804.1835907} sometimes with strong assumptions about the distribution of the attributes \cite{Gao:2010:COE:1835804.1835907}, or separate the context identification from outlier detection \cite{Sanchez:2014:LCS:2618243.2618266}. Other methods detect anomalies without considering any context \cite{ijcai2017-299}.

However, real world networks representing complex economic, social, or biological systems are characterized by modular, multiscalar, and often hierarchical structures \cite{BOCCALETTI2006}. 
Taking into account the node attributes spanning across the \textit{multi-scale} structure of networks to correctly define the contexts of anomalies is therefore crucial. In addition, anomalous nodes might emerge within a specific context and later disappear in different contexts. Hence,  anomaly detection is linked naturally to the problem of community detection, consisting of finding clusters of nodes that are densely linked together compared to the rest of the network \cite{Newman2004}.

By performing multiscalar community detection on a network where related nodes are close to each other through edge weights expressing the similarity of nodes' attributes, we discover \textit{all the relevant contexts} for anomalous nodes. In this sense, we consider that a node can be an outlier only at certain scales and not necessarily at all scales. 

Consider the toy example of the co-working network in Fig. \ref{office_example}. If an employee, e.g., $O_2$, works mainly with employees from a different office and not with the employees of his office, it may appear as an anomaly at the scale of offices. But if his office and the office of the employees he is working with, both belong to the same department, this employee will not be an anomaly at the department scale. At an even larger scale however, a company may form a large context with anomalies persisting as global outliers, e.g, $O_3$. Anomalies are scale dependent, hence considering the multiscale nature of real world systems together with the node attributes is essential for the detection of anomalies.

%Consider the toy example of the work relation network in Fig. \ref{office_example}. Here, colleagues in an office may represent a context at a small scale, where local anomalies emerge, e.g., $O_1,O_2,O_3$ might be persons with attributes in disagreement with the attributes of their colleagues within the office. At a larger scale, one may discover departments as new contexts, so previous anomalies may not make sense, i.e., people from a merged office having similar attributes than $O_2$, and at an even larger scale, a company may form a large context with anomalies persisting as global outliers, i.e., $O_3$ in Fig \ref{office_example}. Considering the multiscalar nature of real world systems together with the node attributes, is therefore essential for the detection of anomalies.

In summary, the contributions of this paper are:
\begin{itemize}
 \item We propose a novel algorithm called MADAN (\textbf{M}ulti-scale \textbf{A}nomaly \textbf{D}etection in \textbf{A}ttributed \textbf{N}etworks) providing a principled mechanism to rank and localize outlier nodes within their \textit{context} at \textit{all scales} in a network.
  
 \item We conduct experiments on synthetic and real world benchmarks showing that our method allows to not only recover and rank the so called ground truth anomalies, but also to discover new anomalies jointly with their contexts.
 
 \item Finally, we show that our method is parallelizable and scales to large networks thanks to the Chebychev approximations of the exponential of the graph Laplacian. As side benefit, it also provides a faster methodology for the continuous-time Markov stability framework \cite{7010026} for community detection.

\end{itemize}

\section{Related work}\label{section2}

As the amount of attributed network data has increased during the last decade, anomaly detection on attributed networks has attracted more attention. 
Analyzing static graphs for anomaly detection can be grouped in two general classes \cite{Akoglu:2015:GBA:2757532.2757629}: anomalies on plain (unlabeled) graphs and anomalies on attributed networks. The former one relates to detecting nodes having anomalous connectivity features in the network without exploiting any node attribute information \cite{10.1007/978-3-642-13672-6_40}, i.e. isolated or bridge nodes. 
The second one, into which our method falls, aims to detect anomalous nodes regarding the attributes of nodes within a given local context, i.e community of a node or attribute subspace.
CODA \cite{Gao:2010:COE:1835804.1835907} is a community based algorithm which uses Hidden Markov Random Fields to characterize both data and links simultaneously turning the outlier detection in an inference problem.
Consub+DisOut \cite{Sanchez:2014:LCS:2618243.2618266} is a two step method performing statistical selection of congruent subspace defined by subset of node attributes and subsequently rank anomalous nodes introducing a distance based outlier model.
ALAD \cite{ijcai2017-325} is a method to retrieve outlier nodes exploiting network structure and node attribute information jointly considered in a non-negative matrix factorization framework.
The RADAR \cite{ijcai2017-299} approach propose a learning framework to identify anomalies via residual analysis exploring the coherence between attribute information and the network information.
AMEN \cite{Perozzi:2018:DCA:3178544.3139241} considers ego-networks to characterize anomalous neighbors in attributed networks. More recent deep learning approaches \cite{Ding:2019:IAD:3289600.3290964} introduce an interactive approach incorporating the feedback form the end user. Unlike the previous methods, our MADAN approach allows to spot anomalous nodes across \textit{all scales} of the network uncovering the relevant scales spanned by the nodes attributes and the graph structure.

\section{Problem statement and framework}
\subsubsection{Creating a weighted graph}
\label{sec:weighting}

Considering a graph  $G=(V,E)$ composed by a set $V$ with $|V|=N$ nodes or vertices, and a set of edges $E$. For simplicity, we will restrict this work to undirected, connected, simple graphs. Considering multidimensional node attributes, we associate to each vertex $u \in V$ a $d$-dimensional vector $\bm{f}(u)$ = $\langle \bm{f}_1(u), \ldots, \bm{f}_d(u) \rangle $ where $\bm{f}_k(u) \in \mathbb{R}$ represents the $k-$th attribute of the vertex $u$. 

Our problem is, informally speaking, to discover anomalous nodes and the context with respect to which they are anomalous, for all possible ranges of context size.

%We first present some prerequisites in graph signal processing and Markov stability framework.

In order to do so, we introduce weights to the edges of the graph, expressing the similarity of the nodes' attributes  
linked by the edge.
Here we use the Gaussian weighting function:
\begin{equation} \label{gaussian_kernel}
    w(u,v) =  \begin{cases} 
      \exp{\left(-\frac{\Vert \bm{f}(u) - \bm{f}(v) \Vert ^2}{2\sigma^2}\right)} & \text{if } (u,v) \in E \\
      0 & \text{otherwise}
   \end{cases}
\end{equation}
where $\sigma$ is a scaling parameter. This creates a weighted adjacency matrix $\bm{W}$ over the network, or similarity matrix, where two nodes are similar if they are neighbours with close values of the attributes. This weighting function is commonly used in image processing to represent similarity between pixel intensities \cite{Haralick:1992:CRV:573190}. Here we use it to translate the node attributes into the structure of the network.

\subsubsection{The heat kernel} 
This adjacency matrix allows us to define a heat equation for any graph signal $\bm{x} \in \mathbb{R}^N$, i.e. any function $V \to \mathbb{R}$ assigning a scalar to every node. The heat equation is expressed as

$$
\bm{\dot{x}}(t)= (\bm{W}-\bm{D})\bm{x}(t) = -\bm{L x}(t)
$$

where $\bm{D}$ is the diagonal matrix of  node strengths, defined by $D_{uu}=d_u=\sum_{v} w(u,v)$. The matrix $\bm{L}$ is called the Laplacian of the weighted graph.

This equation, solved by $\bm{x}(t)=e^{-t\bm{L}}\bm{x}(0)$, takes its name from a physical analogy, where $\bm{x}_u(t)$ is the `temperature' or 'internal energy' of node $u$ (assuming unit heat capacity for each node), which is being redistributed across the graph following Fourier's heat conduction law along edges with conductivity $w(.,.)$. Importantly, the dynamics preserves the average of the signal. This is seen as the sum of every column of $\bm{L}$ is zero. In the physical analogy, it corresponds to the fact that the total energy on the graph is neither created nor destroyed, only diffused. 

In terms of signal processing \cite{6494675}, the heat kernel $e^{-t\bm{L}}$ is often seen as a smoothing filter acting on the initial signal $\bm{x}(0)$, parametrized by the time $t$. It replaces every entry of $\bm{x}(0)$ with a weighted average of other nodes' signals (with a greater emphasis on neighboring nodes).  In the limit of large $t$, the smoothed signal is just constant over the network. 

When the graph is a discrete line with constant weights (modelling for example a discrete unidimensional space), the filter is translation-invariant. This means that for any two Kronecker delta signals $\bm{\delta}_u$ (taking unit value on node $u$ and zero value elsewhere) or $\bm{\delta}_v$ (for node $v$),  the corresponding smoothed signals  $e^{-t\bm{L}} \bm{\delta}_u$ and $e^{-t\bm{L}} \bm{\delta}_v$ are translations of one another, and therefore have the same shape. This translation-invariance, typical in `standard' signal processing, disappears in arbitrary networks, where  $e^{-t\bm{L}} \bm{\delta}_u$ will strongly depend on the structure and weights of the network around $u$. This property allows us to characterize the structure of the graph through the properties of its kernel. We use this fact first to detect the anomalies and then to detect the contexts in which these anomalies are relevant.
\subsubsection{Finding the  anomalous nodes}
The \emph{concentration} of a node $u \in V$ at scale $t$ is defined by the $L_2-$norm of the filtered $\bm{\delta}_u$ signal with a heat kernel at time $t$ \cite{perraudin_ricaud_shuman_vandergheynst_2018}:
\begin{equation}\label{concentration}
c_u(t) = \Vert e^{-t\bm{L}}\bm{\delta}_u \Vert_2.
\end{equation}

This measure is useful because the filter preserves the sum (or mean) of the signal, and therefore the Kronecker delta has maximum norm of one while the completely smoothed signal has norm $1/N$, which is the smallest possible norm for a signal of sum one spread over $N$ nodes.     

A node with high concentration at time $t$ tends to be poorly connected with its neighborhood, through few edges or edges of low weight. The time $t$ selects the size of the neighborhood of reference, in a way that is discussed below. 

With the weighting proposed in Eq. \ref{gaussian_kernel}, a high-concentration node indicates a node that is very dissimilar in its attributes with its neighbors.
In this way, we use the concentration as way to quantify the degree of deviation of a given node with respect to its context at a given time scale and provides a scoring for ranking potential outliers. 

To identify outliers, we use the following standard thresholding rule used in general outlier detection methods \cite{trove.nla.gov.au/work/10589055} which works well in our experiments.
However, other criteria for identifying from the concentration distribution could be developed. 

Let $c(t) = [c_1(t), \ldots , c_N(t)]$ be the overall graph concentration. A node $u$ is considered as \emph{anomalous} at a time scale $t$ if the following thresholding condition holds:
\begin{equation} \label{threshold}
c_u(t) \geq \bar{c}(t) + 2 s(c(t))  \\
\end{equation}

with $\bar{c}(t)$ as the average concentration and $s(c(t))$ its standard deviation across nodes.

\subsubsection{Finding the contexts} \label{sec:contexts}

We now make sense of the choice of $t$, in that it selects the context with respect to which a node is anomalous. In a large $t$ limit, a delta signal  $\bm{\delta}_u$ will be smoothed to a constant over the whole graph, as the heat kernel will have the time to mingle the signal of all nodes together. A node that stands as an outlier for a large $t$ is therefore an outlier globally, with respect to the whole graph.
Conversely, a small $t$ only allows heat diffusion with immediate neighbors, and thus a small $t$ outlier is anomalously dissimilar to a small context. 

More generally, we consider that a set of nodes $S$ is a suitable context for all potentially anomalous nodes lying in it, if this set is relatively poorly connected to the rest of the network, in a manner akin to community detection. Here, `poorly connected' is defined in reference to a given time scale $t$, in that the `internal energy' contained in $S$ essentially remains in $S$ within time $t$.  Consider $\bm{h}_S$ the characteristic signal of $S$, i.e. the node signal taking unit values  in $S$ and zero values outside. The initial total energy of $S$ is $|\bm{h}_S|_1$, the number of nodes in $S$. After smoothing, the energy remaining in $S$ in excess of the energy $\frac{1}{N}|\bm{h}_S|_1$ that will remain at $t=\infty$  is: 

\begin{equation}
r(t;\bm{h}_S) = \bm{h}_S^T e^{-t\bm{L}} \bm{h}_S - \frac{1}{N}|\bm{h}_S|_1,
\label{markov_stability2}
\end{equation}
which we want to be as high as possible.

As we want to be able to provide a context for potentially each node of the graph, we look for a partition of the nodes, encoded by its $N \times K$ characteristic matrix $\bm{H}$, where $K$ is the number of sets and every column $\bm{h}_1, \ldots, \bm{h}_K$ is the characteristic vector of a set of nodes. An optimal partition into contexts is given by the matrix $\bm{H}$ maximizing:

\begin{equation}
r(t;\bm{H}) =  \sum_i \left( \bm{h}_i^T e^{-t\bm{L}} \bm{h}_i - \frac{1}{N}|\bm{h}_i|_1\right).
\label{markov_stability2}
\end{equation}

This is essentially a particular case of the Markov stability \cite{Delvenne12755,7010026}, designed as a general framework for multi-scale community detection. It has the expected behaviour to provide a fine partition of small sets of nodes for low $t$, and a few large sets for large $t$. As some nodes could be assigned to their own isolated context, in practice, we join them with the closest one.

\begin{figure*}[t!]
\centering
\hspace*{-0.3cm}
\includegraphics[height=6.7cm]{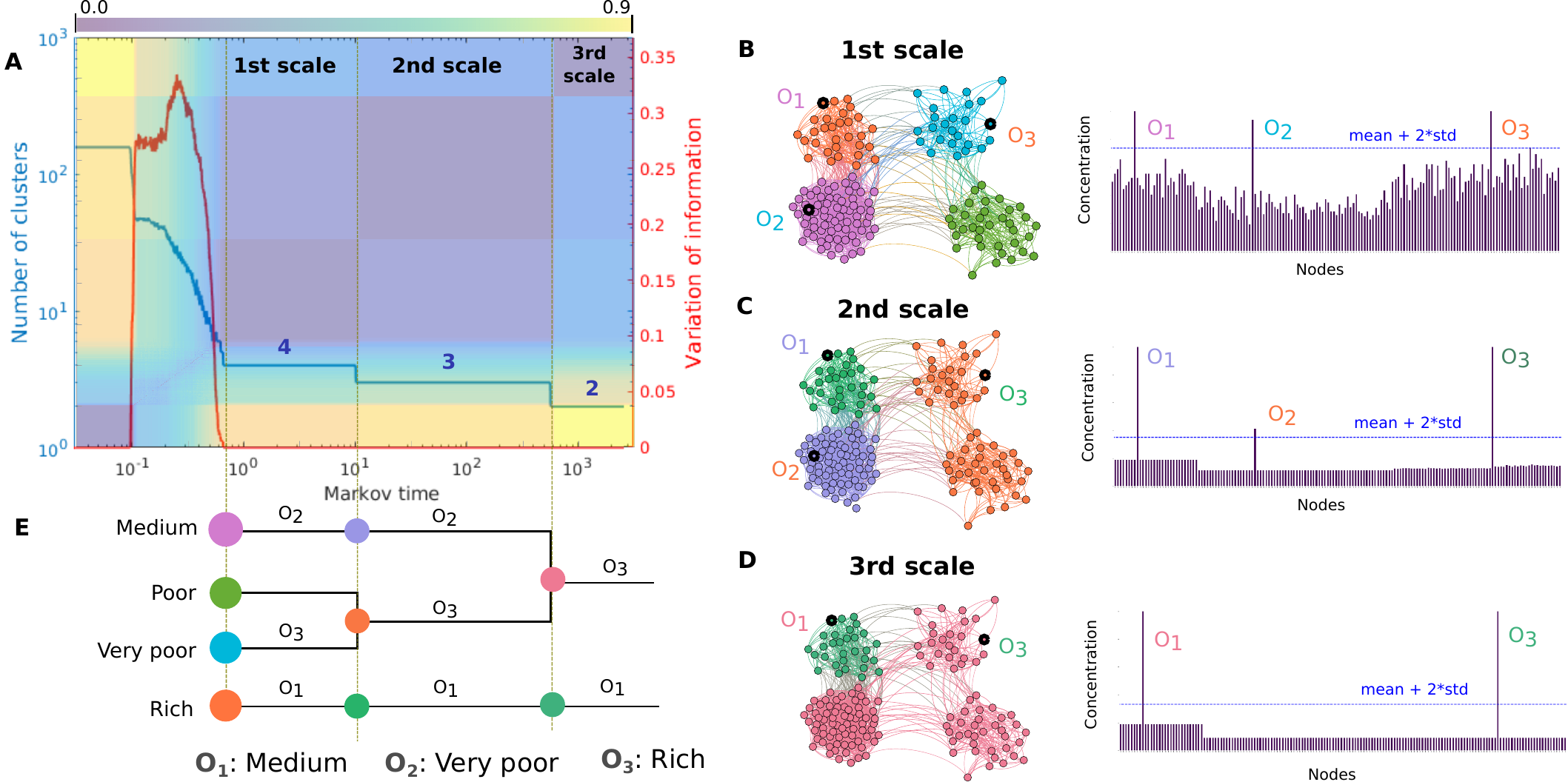}
\caption{Synthetic social network of 160 nodes with a planted partition of four communities. Each node (representing a person) has a scalar attribute representing the income of a person, partitioning the network in four clusters labeled as rich, medium, poor, and very poor. Three outlier nodes are injected: $O_1$ is a medium income person within the rich cluster, $O_2$ a very poor person within the cluster of medium income people and $O_3$ is a rich person in the very poor cluster. \textbf{(A)} We perform a scanning of relevant contexts along the attribute subspace: for each time $t$ we find the number of clusters (blue curve) and compute the average variation of information (VI$(t)$)) of the retrieved partitions (red curve). The VI$(t)$ is obtained from 100 runs of the Louvain algorithm. The background represents the VI$(t,t')$ between optimized partitions across times. As can be seen in \textbf{(A)}, starting at $t\approx 0.65$, the partitions of $4$, $3$ and $2$ clusters are found as persistent according with the low VI$(t)$.
Subsequently, we compute the concentration over all nodes of the graph (bar plots) with Eq. (\ref{concentration}). 
\textbf{(B)} At the 1st scale, the concentration of nodes at $t=1.5$ reveals three local outliers with their context given by the colored clusters. \textbf{(C)} At the 2nd scale, at $t=10.5$ the context of $O_3$ enlarges whereas the concentration of $O_2$ decreases. \textbf{(D)} At the 3rd scale given by $t=578$, $O_2$ is not an outlier and the context for $O_3$ enlarges more. \textbf{(E)} The dendrogram shows the hierarchy of partitions and the outlier nodes at each scale.}
\label{toy_example}
\end{figure*}

Although it is known to be NP-hard to optimize the maximization of Eq. (\ref{markov_stability2}) \cite{4358966} it can be recoded \cite{7010026} into a modularity maximization problem for which the Louvain algorithm \cite{Blondel_2008} is known to provide good results in practice. For each time step, we run the Louvain algorithm $100$ times with different random initializations and use Eq. \ref{markov_stability2} as a scoring function to rank partitions $\bm{H}$. In order to assess the robustness of
the retrieved partition at a given time scale, we follow the methodology of \cite{Delvenne2013} and compute the variation of information among the ensemble of found partitions for a given time. To be more precise, the normalized variation of information of two partitions $P_1$ and $P_2$ reads:

\begin{equation}
    VI(P_1,P_2)(t) = \frac{H(P_1|P_2) + H(P_2|P_1)}{\log(N)}
\end{equation}
where $H(P_1|P_2)$ is the conditional entropy of $P_1$ given $P_2$, i.e the  additional information needed to describe $P_1$ when $P_2$ is known assuming a uniform probability on the nodes.

To simplify the notation we will write $VI(P_1,P_2)(t)$ as $VI(t)$. 
To find the range of scales relevant for a given partition, we also compute the variation of information between the ensemble of partitions found between each pair of times $VI(t,t')$.
Relevant partitions $\bm{H}$ are finally identified by times $t, t'$ where $VI(t)$ shows a dip and $VI(t,t')$ shows a continuous plateau \cite{7010026}.

\subsubsection{Summary of our approach}
Our multi-scale anomaly detection method (MADAN) is summarized and illustrated in Fig. \ref{toy_example}. Roughly speaking, we first scan the network to find the most relevant scales and finding the contexts, as described in the previous section. For the times that are identified by this context-finding methodology, we simultaneously detect the outliers satisfying Eq. \ref{threshold}. 

\section{Experiments}\label{section4}
We evaluate the performance of our method in detecting anomalous nodes in synthetic and real life attributed networks. MADAN\footnote{MADAN Python code can be found at https://github.com/leoguti85/MADAN} is compared against state-of-the-art and baseline methods.

\subsection{Experimental setup}
\subsubsection{Synthetic dataset.}
We generate a series of undirected synthetic networks with ground truth community structure, adding node attributes and injecting node anomalies in different hierarchies. 

\begin{figure}[t!]
\centering
\hspace*{-0.6cm}
\includegraphics[height=3.4cm]{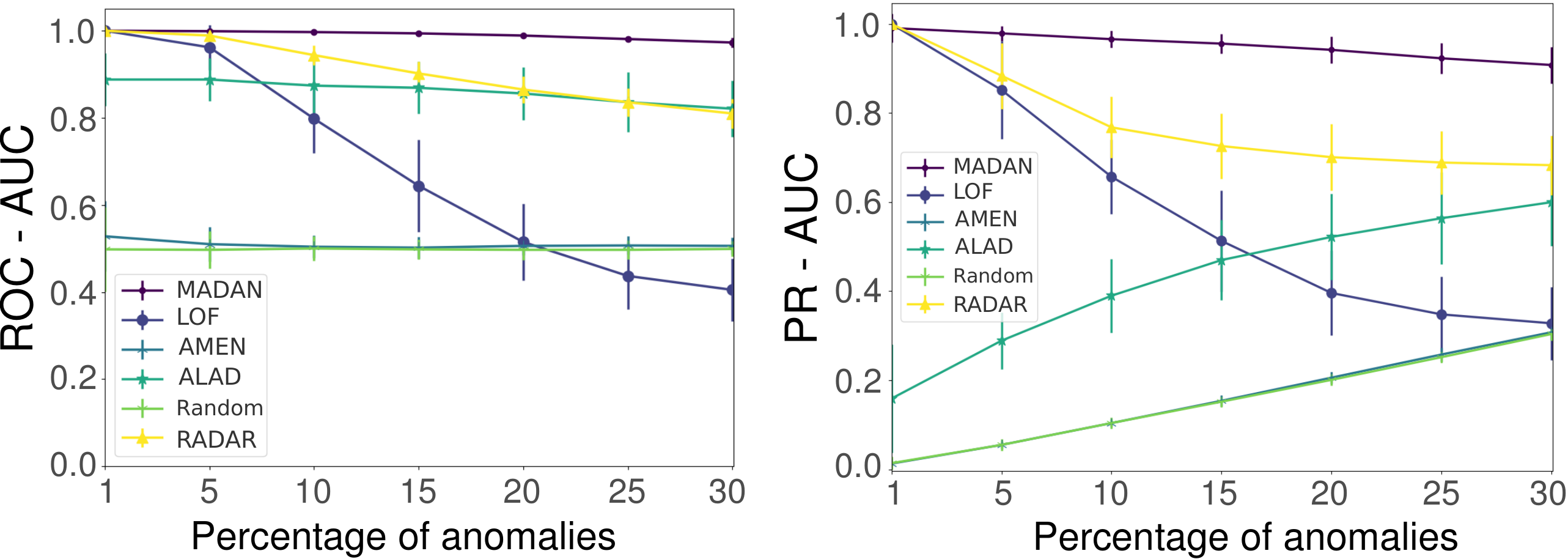}
\caption{Synthetic experiments. Performance on anomaly detection when adding increasing levels of anomalous nodes. Colors correspond to different baseline techniques, including our MADAN approach. We report the mean and standard deviation (vertical lines) over $50$ LFR networks.}% of the ROC-AUC values (\textbf{left}) and the PR-AUC (\textbf{right}).
\label{auc_plots}
\end{figure}

In order to simulate real networks, we generate networks with the Lanchinetti-Fornunato-Radicchi (LFR) benchmark \cite{PhysRevE.80.056117}. This algorithm generates networks with diverse community structures, community sizes and power law degree distributions. Networks with $N=1000$ nodes are generated with the mixing parameter $\mu=0.1$ generating partitions of well defined communities containing between $50$ to $200$ nodes. The average node degree is ranging from $10$ to $100$. 
The size of the communities are taken from a power law distribution with exponent $1$. Similarly, the degree distribution for the nodes is drawn from a power law distribution with exponent equal to $2$.

Nodes attributes are generated as follows: to each node, we associate a vector of attributes of dimension $d=20$. To make the problem more realistic, we introduce heterogeneity in the node attribute values, so that within each community $C \in \{0,1,2,\ldots, c_{max} \}$ nodes attributes are sampled from different probability distributions. 

For any attribute $k$ of a node belonging to a community $C$, its values are drawn from a normal distribution with mean $C$ and standard deviation $\hat{\sigma}$, a uniform distribution in the interval $\left[ C-\hat{\sigma}, C + \hat{\sigma} \right] $ and a logistic distribution with mean $C$ and scale $\hat{\sigma}$.
As nodes within a same cluster are deemed to have similar values, we chose $\hat{\sigma}=0.1$ for all distributions.
In addition, we force some clusters to have similar node attributes, i.e. drawn from the same distributions, in order to introduce attribute hierarchies. 
Thus, the ground truth community structure defines structural clusters, and node attributes on top introduces clusters similarity, i.e., clusters of clusters.
To generate anomalous nodes, a percentage of randomly chosen nodes ($1\%,5\%,10\%,15\%,20\%,25\%,30\%$) is perturbed by replacing $30\%$ of their attributes with attributes from a distinct, randomly chosen cluster. 
The performance of anomaly detection is measured over $50$ random networks, reporting the average and standard deviation of the performance measures.

\subsubsection{Real life datasets.}
We also evaluate our approach on three real life attributed networks, see Table \ref{real_datasets}. They were labeled with the so called ground truth anomalies used as benchmark datasets in the literature.

\begin{table}[h!]
\scriptsize
\caption{Real life datasets}
\centering
\begin{tabular}{c|c|c|c|c}
\hline
\textbf{DATASET} & \textbf{\textit{\# nodes}}& \textbf{\textit{\# edges}}& \textbf{\textit{\# node attr}} & \textbf{\textit{\# anomalies}} \\
\hline
Disney & 124 & 334 & 28 & 6 \\

Books & 1418 & 3695 & 28 & 28 \\

Enron & 13533 & 176987 & 20 & 5 \\
\hline
\end{tabular}
\label{real_datasets}
\end{table}

The Disney and Books datasets \cite{6547453} are co-purchase networks extracted from Amazon. They contain $28$ attributes per node describing properties about online items, i.e., rating, selling price, etc. 
The ground truth anomalies for Disney DVDs movies were tagged manually by high school students in Germany.
Ground truth anomalies were defined as follow \cite{6547453}.
A partition of the network was obtained after applying the Louvain algorithm. Students were then asked to tag anomalous nodes with a high deviation of the attribute values within each cluster. Outliers nodes were defined as the ones tagged as anomalous by at least $50\%$ of the students.
In the Book dataset, ground truth anomalies were defined as nodes having the tag \textit{amazonfail} \cite{6547453}.

Enron \cite{Metsis06spamfiltering} is a communication network with edges indicating email transmission between people. Each node contains $20$ attributes describing metadata of the message i.e., content length, number of recipients, etc. This dataset has been extensively used as benchmark for spam detection \cite{Metsis06spamfiltering}. Spammers were labeled as ground truth for anomaly detection \cite{6547453}. In all cases, we set the parameter $\sigma$ in Eq. \ref{gaussian_kernel} as the standard deviation of the distribution of pairwise distances between node attributes.

\subsubsection{Evaluation metric.}
Following the evaluation setup of \cite{ijcai2017-299,6547453}, 
we assess the anomaly detection performance with two well known metrics for evaluation of anomaly detection systems \cite{Aggarwal:2013:OA:2436823}: the area under the receiver operating characteristic curve (\textbf{ROC-AUC}) and the area under the precision/recall curve (\textbf{PR-AUC}). The former allows to quantify the trade-off between true positive rate (tpr) and false positive rate (fpr) across different thresholds. 
The tpr is defined as the detection rate, i.e. the rate of true anomalous nodes correctly identified as anomalous, whereas the fpr is the false alarm rate, i.e. rate of normal nodes identified as anomalous. The second metric quantifies the trade-off between the precision, i.e., the predictive power of the method and the recall, defined as the ratio between true positive over true positives plus false negatives. In both cases, the more the AUC approaches to $1$ the better the performance of the method.

\subsubsection{Baseline methods}
We compare our MADAN approach with five baseline methods of the literature having their source code available: Local Outlier Factor (LOF) \cite{Breunig:2000:LID:335191.335388}, Accelerated Local Anomaly Detector (ALAD) \cite{ijcai2017-325}, Anomaly Mining of Entity Neighborhoods (AMEN) \cite{Perozzi:2018:DCA:3178544.3139241}, Residual Analysis for Anomaly Detection (RADAR) \cite{ijcai2017-299} and a simple Random classifier (Random) for anomaly detection, i.e., assigning random ranking scores to each node. We did not experiment with CODA \cite{Gao:2010:COE:1835804.1835907} and Consub+DisOut (C+D) \cite{Sanchez:2014:LCS:2618243.2618266} on our synthetic dataset because the authors did not provide their scripts. However, we compare against them on the real life benchmarks by taking their performance values from their papers. Each method output an anomaly score or a ranking of nodes according with their anomaly level. We compute the ROC and PR curves and report their AUC score.

\subsection{Performance evaluation}
\subsubsection{Synthetic results.}
Results on synthetic attributed networks are shown in Fig. \ref{auc_plots}. As can be seen, our MADAN approach outperforms the baseline methods in all cases. Here the scale $t$ was chosen as the one giving the best performance. Indeed our approach can deal with hierarchical composition of attributes across communities together with heterogeneous distribution of node attributes. Injecting a few amount of anomalies, i.e., $1\%$ of perturbed nodes, MADAN is comparable with RADAR and LOF in both ROC-AUC and PR-AUC metrics. However, when the number of anomalies increases, LOF performance fall down because it does not exploit the graph structure. It can also be seen that AMEN performance is precarious comparable to a random classifier because this method was designed to detect anomalous nodes on ego-networks discovering anomalous neighborhoods. MADAN also shows the lowest variance across identifying outliers across different random networks.

\subsubsection{Real life results.}
Results on real life datasets are shown in Table \ref{results_real}. 
We see that our approach (MADAN) achieves the best performance compared to other baseline methods on the Disney and Books networks while C+D shows the best AUC on the Enron dataset.
The advantage of our method that it allows us to find the right scale for which the anomalous nodes are well defined regarding the nodes attributes of the network. In general it performs better, or at least with comparable AUC, than the baseline methods, in particular compared to LOF, which ignores the underlying structure of the network, and AMEN, which focuses only anomalous neighbors. Results of precision, recall and F1\footnote{The F1-score is weighted by support (the number of true instances for each label), accounting for label imbalance.} scores (thresholded with Eq. \ref{threshold}) in Table \ref{results2_real} highlight the performance of our method, except on Enron dataset. This shows that our method is not well suited for spammer detection.

\begin{table}[t]
\scriptsize
\caption{Results real life datasets (ROC-AUC)}
\centering
\begin{tabular}{ccccccc}
\hline
\textbf{DATASET} & \textbf{LOF}& \textbf{C+D}& \textbf{RADAR} & \textbf{ALAD}& \textbf{AMEN} & \textbf{MADAN}\\
\hline
Disney & 0.61 & 0.81 & 0.87 & 0.70 & 0.52 & \textbf{0.93}\\
Books  & 0.49 & 0.60 & 0.58 & 0.43 & 0.47 & \textbf{0.68} \\
Enron  & 0.44 & \textbf{0.74} & 0.65  & 0.72 & 0.47 & 0.66\\
\hline
\end{tabular}
\label{results_real}
\end{table}

It is worth mentioning that, when assessing the performance of anomaly detection algorithms on empirical datasets, one should keep in mind that the set of manually annotated, so-called, `ground truth' anomalies does not represent an objective truth as different annotators may label different nodes as anomalous, choosing different criteria to make their decisions. 
We believe that our method, that allows to measure the degree of anomaly of each node as a function of the size of its attribute context, offers a framework to explore systematically anomalies in a principled way. We will present later an example of such anomaly exploration in a real world network..

\subsection{Computational issues}
The computational bottleneck of our approach relies in the computation of the exponential of the Laplacian for large graphs (more than 8000 nodes)
used for computing the node concentration (Eq. \ref{concentration}), and finding the optimal context for anomalous nodes (Eq. \ref{markov_stability2}).
Previously, \cite{HAMMOND2011129} introduce an efficient way to approximate kernel evaluations of general filters employing Chebychev polynomials approximation. Here\footnote{All computations were done on a standard computer Intel(R) Core(TM) i7-4790CPU, 3.60GHzI with 16G of RAM.} we propose a way to compute the heat kernel evaluation at a given scale, by approximating the entire exponential of the Laplacian as:

\begin{equation}
    \bm{e}^{-t\bm{L}} = [\bm{e}^{-t\bm{L}}\bm{\delta}_1, \ldots, \bm{e}^{-t\bm{L}}\bm{\delta}_N  ]
\end{equation}

where each column component is approximated with a Chebychev polynomial, being easily parallelized across columns. The error of the approximation is controlled by the degree of the Chebychev polynomial $m$. The larger the $m$ the better the approximation with a cost of a longer running time. We chose $m=30$ as in \cite{HAMMOND2011129}. 

\begin{table}[t]
\scriptsize
\caption{Results real life datasets (PR/RC/F1-score)}
\centering
\begin{tabular}{clcccccc}
\hline
\textbf{} & \textbf{DATASET} & \textbf{LOF}& \textbf{RADAR} & \textbf{ALAD}& \textbf{AMEN} & \textbf{MADAN}\\
\hline
&Disney & 0.55 & 0.47 & 0.53 & 0.48 & \textbf{0.82}\\
&Books  & 0.50 & 0.50 & 0.50 & 0.50 & \textbf{0.52} \\
\rot{\rlap{Precision}} 
&Enron  & 0.49 & 0.50 & 0.50 & 0.50 & 0.50\\
\hline
&Disney & 0.62 & 0.50 & 0.67 & 0.41 & \textbf{0.82}\\
&Books  & 0.50 & 0.52 & 0.46 & 0.44 & \textbf{0.60} \\
\rot{\rlap{Recall}} 
&Enron  & 0.44 & 0.55 & 0.55 & 0.45 & \textbf{0.60}\\
\hline
&Disney & 0.57 & 0.48 & 0.41 & 0.35 & \textbf{0.83}\\
&Books  & 0.48 & 0.09 & 0.34 & 0.34 & \textbf{0.53} \\
\rot{\rlap{F1-score}} 
&Enron  & \textbf{0.47} & 0.10 & 0.33 & 0.33 & 0.17\\
\hline
\end{tabular}
\label{results2_real}
\end{table}

\begin{figure}[h!]
\centering
\includegraphics[height=3.2cm]{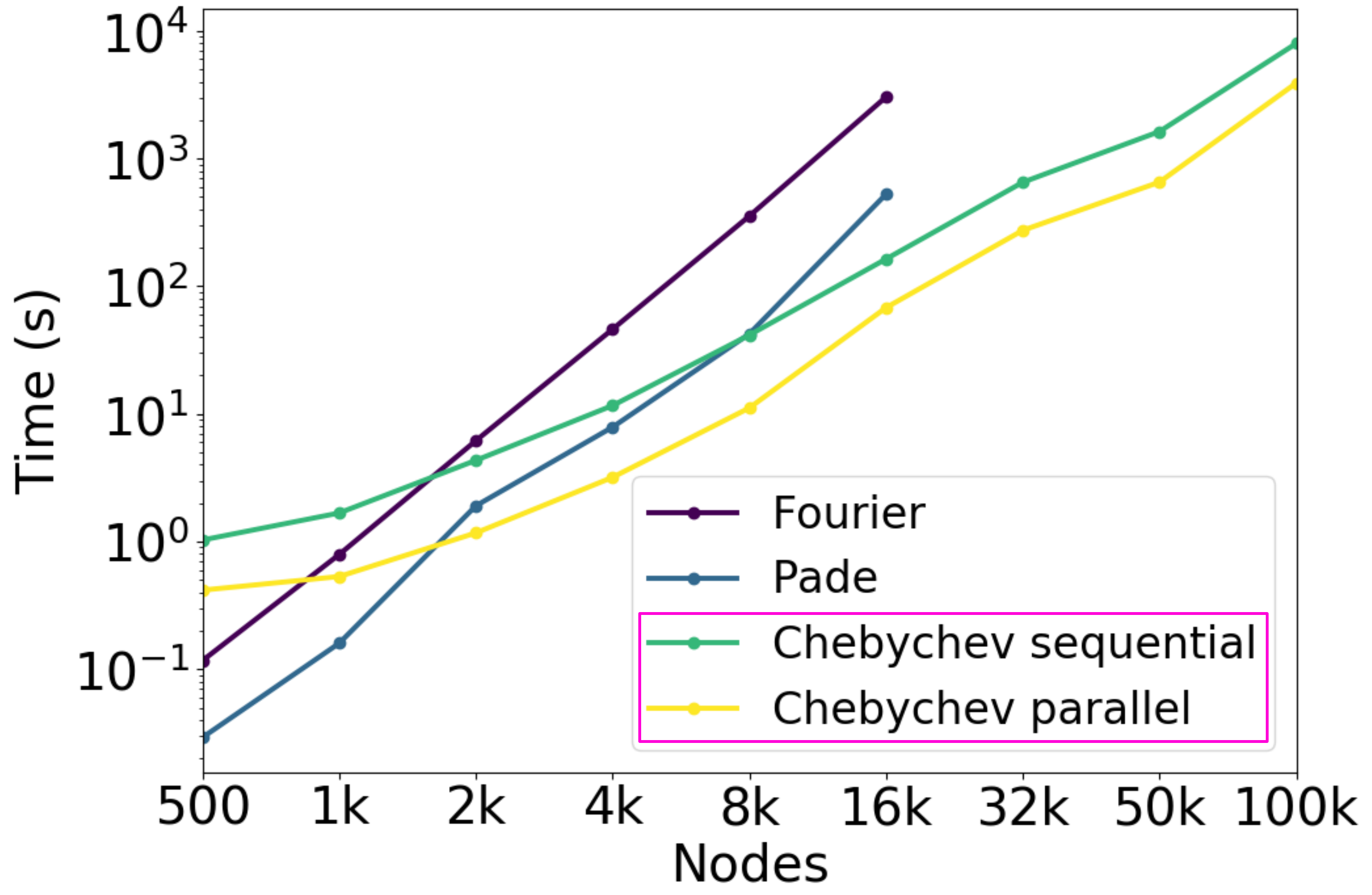}
\caption{Running time for the computation of the exponential of the graph Laplacian for $t=1$ varying the network size. Our parallel setting run across 8 cores.}
\label{runningtime}
\end{figure}

We test the scalability of this approach in the computation of the exponential of the Laplacian, compared with the classical Fourier approach, i.e., Cholesky decomposition and Pad\'{e} approximations. We generate Barab\'{a}si-Albert networks with power-law degree distribution of varying number of nodes $N$ and number of edges 
equal to $3N$. Results are depicted in Fig. \ref{runningtime}.

\begin{figure*}[h!]
\centering
\includegraphics[height=9.8cm]{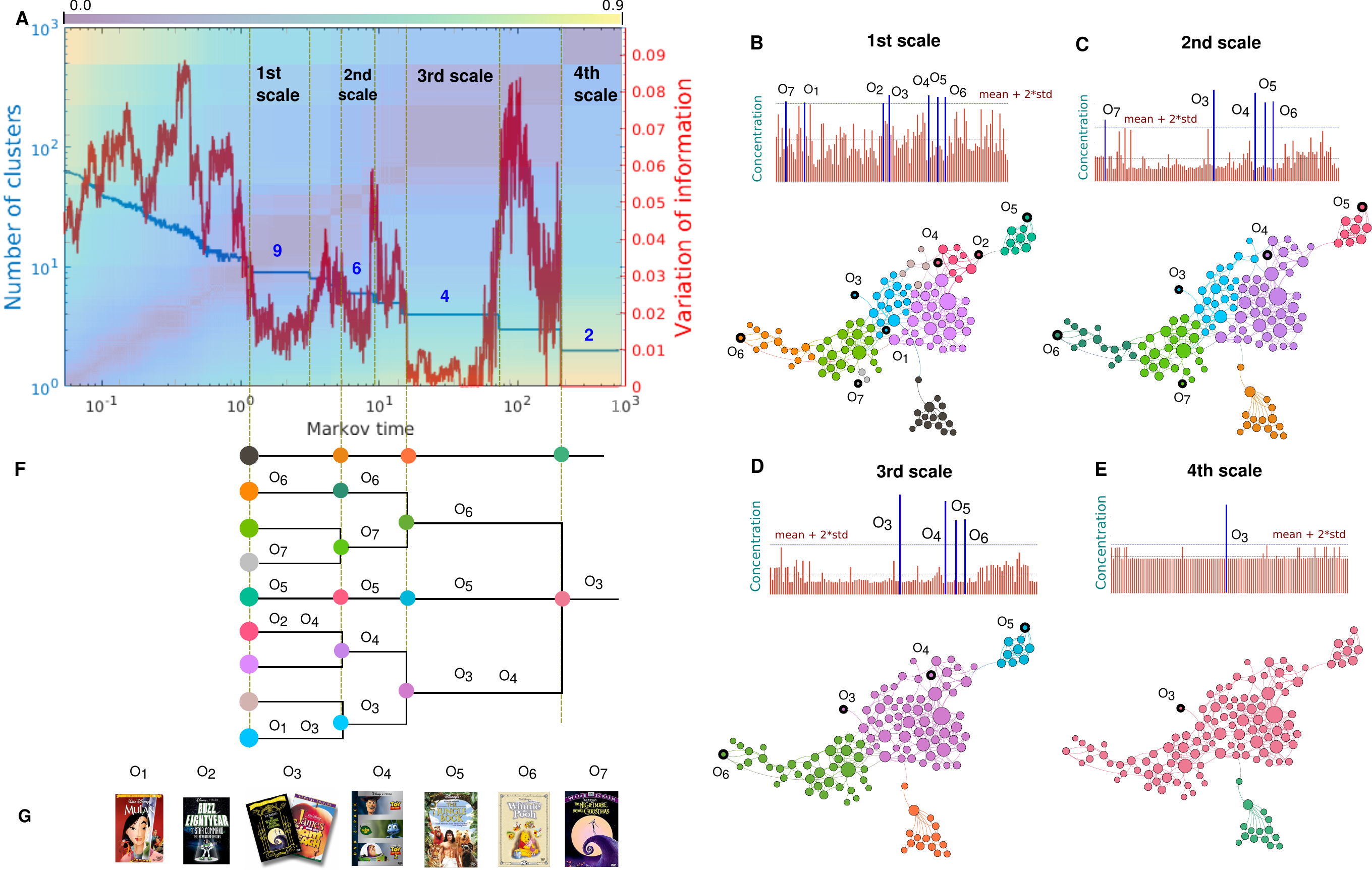}
\caption{\textbf{(A)} Number of clusters (blue curve) found by maximizing Eq. \ref{markov_stability2}, at each time step, the variation of information $VI(t)$ (red curve) between the ensemble of optimal partitions at each time and the variation of information ($VI(t,t')$) between optimal partitions across times (background contour plot).
Relevant partitions are determined by dips of $VI(t)$ and extended plateaus of $V(t,t')$.
Visualization of four robust partitions and the node heat concentration (bar plots), indicating outlier nodes (blue bars) when evaluating the node concentration at \textbf{(B)} $t=1.18$, \textbf{(C)} $t=7.25$, \textbf{(D)} $t=14.89$ and \textbf{(E)} $t=204.8$. In all bar plots, the upper horizontal line (blue) indicates the detection threshold defined in Eq. \ref{threshold}. \textbf{(F)} Dendrogram showing the hierarchy of clusters at each time with the contextual outliers. \textbf{(G)} Anomalous Disney DVD movies.}
\label{disney_plot}
\end{figure*}

\section{Multi-scale anomaly detection: a case study}\label{section5}
We consider the Disney network as a case study. This dataset provides manually labeled ground-truth anomalies \cite{6547453} and has been largely used as benchmark for outlier detection in networks.
Here, we use \textit{MinPricePrivateSeller} and \textit{Number\_of\_reviews} as node attributes. 

As is shown in Fig. \ref{disney_plot}, we start scanning to look at relevant scales taken into account both the node metadata and the graph structure.
For each intrinsic scales found in Fig. \ref{disney_plot}A, we display on the network the corresponding optimal partition (scales) in the right-hand side of Fig. \ref{disney_plot}.

The contour plot in Fig. \ref{disney_plot}A shows the variation of information $VI(t,t')$ between the sets of optimal partitions at time $t$ and $t'$ and reveals blocks of low VI (dark blue blocks) corresponding to the different scales uncovering different contexts.
The number of clusters at each scale (blue curve) displays plateaus corresponding to the blocks in $VI(t,t')$.
The scales are also visible as dips in the $VI(t)$ among the set of partitions found by iterating the Louvain algorithm at a given time scale (red curve). We identify four relevant scales with $9$, $6$, $4$ and $2$ number of clusters (or contexts) respectively. Simultaneously, the concentration on each node Eq. \ref{concentration} is evaluated at the same $t$, spotting anomalies at the given scale and context.

As shown in the right hand side of Fig. \ref{disney_plot}, (bar plots in \textit{B,C,D,E}), outliers nodes at a given scale are detected as the ones having a concentration of energy more than two standard deviation above the mean concentration (Eq. \ref{threshold}).

At the \textbf{1st scale} (the finest one) the Fig. \ref{disney_plot}B shows the partitioning of the network in $9$ clusters. Node $O_5$ (\textit{The jungle book}) appears as an anomalous node within this context, that corresponds to all the \textit{read along} movies, having both high selling price and number of reviews.
An interesting case is found in $O_3$ (\textit{The Nightmare Before Christmas / James and the Giant Peach}) which has low reviews but is a very expensive article because it is the only one in its context consisting in a two DVD pack. 
The cluster containing $O_2$ and $O_4$ corresponds to Pixar movies, where again $O_4$ is a three pack DVD promotion (\textit{Toy story 1-2} and \textit{A Bug's Life}), whereas $O_2$ corresponds to the poorly reviewed but very expensive \textit{Buzz Lightyear of Star Command: The Adventure Begins} film.
The cluster containing $O_6$ consists in  many \textit{Winnie-The-Pooh} related movies with some independent $\textit{Tigger}$ sagas. The anomalous node $O_6$ (\textit{The Many Adventures of Winnie the Pooh}) is the most expensive film with the highest rate of reviews in this category. The three-nodes cluster containing $O_7$ (\textit{The Nightmare Before Christmas}) flagged $O_7$ as an anomaly because its selling price is $3$ and $6$ times higher than the two other nodes.

Going up in the hierarchy of clusters allows us to detect anomalous nodes within coarser contexts and discarding more local outliers. 
As can be seen in the \textbf{3rd scale}, the Fig. \ref{disney_plot}D, the outliers remaining within the pink cluster, $O_3$ and $O_4$, correspond to the two packs DVD promotions with very high prices, whereas more local anomalies disappear at this scale, e.g., $O_7$.

Finally we see in Fig. \ref{disney_plot}E that at the \textbf{4th scale}, $O_3$ (\textit{The Nightmare Before Christmas / James and the Giant Peach}) remains as a global outlier within the coarser pink cluster, being a bad connected node with high attribute values.

\section{Conclusions} \label{section6}
In this work we introduce an effective algorithm (MADAN) to perform multi-scale anomaly detection on attributed networks. 
Anomalous nodes are characterized as those remaining highly concentrated after smoothing unit impulse signals around each vertex of the graph. Using the heat kernel as filtering operator allows us to exploit the link with the Markov stability to find the context for outlier nodes at \textit{all relevant scales} of the network. Extensive empirical studies on synthetic and real world benchmarks demonstrated the superiority of MADAN over many state of the art approaches.
In addition, we show that our method is highly efficient, being easily parallelized scaling with networks with up to at least 100k nodes.

\section{Acknowledgments}
This work was supported by the Flagship European Research Area Network (FLAG-ERA) Joint Transnational Call ``FuturICT  2.0'' and the Swiss National Science Foundation Fellowship P300P2\textunderscore 177793 (A.B) which are gratefully acknowledged as well as Leto Peel for helpful suggestions.

\fontsize{9.5pt}{10.5pt}
\selectfont
\bibliographystyle{aaai}
\bibliography{references} 

\end{document}